\documentclass[a4paper,fleqn,usenatbib]{mnras} 

\usepackage{newtxtext,newtxmath}

\usepackage[T1]{fontenc}
\usepackage{ae,aecompl}

\usepackage{graphicx}	
\usepackage{amsmath}	
\usepackage{amssymb}	
\usepackage{booktabs}
\usepackage{color}
\usepackage{pdflscape}
\usepackage{subfig}

\title[Towards Machine-assisted Meta-Studies]{Towards Machine-assisted Meta-Studies: The Hubble Constant}

\author[T. Crossland, et al.]{
Tom Crossland,$^{1,2}$\thanks{E-mail:~\texttt{t.crossland.17@ucl.ac.uk}}
Pontus Stenetorp,$^{2}$
Sebastian Riedel,$^{2}$
Daisuke Kawata,$^{1}$
\newauthor{}
Thomas D. Kitching,$^{1}$
and Rupert A. C. Croft$^{3}$
\\
$^{1}$Mullard Space Science Laboratory, University College London, Holmbury St. Mary, Dorking, Surrey RH5 6NT, United Kingdom\\
$^{2}$Department of Computer Science, University College London, Gower Street, London WC1E 6BT, United Kingdom\\
$^{3}$McWilliams Center for Cosmology, Department of Physics, Carnegie Mellon University, 5000 Forbes Avenue, Pittsburgh, PA 15213, USA
}

\date{Accepted XXX. Received YYY; in original form ZZZ}

\pubyear{2018}

\begin{document}
\label{firstpage}
\pagerange{\pageref{firstpage}--\pageref{lastpage}}
\maketitle

\begin{abstract}

We present an approach for automatic extraction of measured values from the astrophysical literature, using the Hubble constant for our pilot study. Our rules-based model -- a classical technique in natural language processing -- has successfully extracted 298 measurements of the Hubble constant, with uncertainties, from the 208,541 available arXiv astrophysics papers. We have also created an artificial neural network classifier to identify papers in arXiv which report novel measurements. From the analysis of our results we find that reporting measurements with uncertainties and the correct units is critical information when distinguishing novel measurements in free text. Our results correctly highlight the current tension for measurements of the Hubble constant and recover the $3.5\sigma$ discrepancy -- demonstrating that the tool presented in this paper is useful for meta-studies of astrophysical measurements from a large number of publications.

\end{abstract}




\section{Introduction}


The increase in publication output of the scientific community has, in recent years, surpassed the level at which most academics can stay up to date. Even if one chooses a narrow focus, more papers are published each month than can be practically read by any one individual in the given time. Further, if one wishes to make a formal study of the value of a given parameter, across the multiple publications in which such measurements are reported, this problem is compounded by the need to find the various publications in the first place. The results of such studies are not only interesting as observations on the state of the community and its collective knowledge, but are also very useful for determining consensus (or lack thereof) and highlighting issues which merit further study. Structured analysis of the body of existing measurements can be used to refine simulations and models, and also to motivate directions in research if discrepancies or consensus can be found. 

For example, a series of papers from \citet{deGrijs+2014AJ....148...17D,deGrijs+2015AJ....149..179D,deGrijs+Bono2016ApJS..227....5D,deGrijs+Bono2017ApJS..232...22D} discussed publication bias in measurements of the distances to the Local Group Galaxies, and Galactic rotation properties. Similarly, \citet{Croft+Dailey2011arXiv1112.3108C} compiled measurements of cosmological parameters between 1990 and 2010, and noted a confirmation bias when comparing the scatter between the resulting measurements, given reported uncertainties.
\citet{Licquia+Newman2015ApJ...806...96L} compiled measurements of Milky Way properties from the literature, and performed a sophisticated statistical analysis on the resulting data. Regarding the Hubble constant, John Huchra undertook to compile published measurements of the Hubble constant between 1996 and 2010, and his results\footnote{\url{https://www.cfa.harvard.edu/~dfabricant/huchra/hubble/index.htm}} have been used as a basis for many meta-studies, such as \citet{Gott2001ApJ...549....1G} and \citet{Zhang2018PASP..130h4502Z}. Additionally, a review of the measurements of the Hubble constant is given by \citet{Freedman+Madore2010ARA&A..48..673F}.
%

However, conducting such meta-studies is time-consuming, and often laborious -- factors which themselves can lead to human and clerical errors in the collating of information. But with this growth in publication output there is a growing corpus of literature -- especially in the physical sciences -- which, along with recent advances in machine learning and natural language processing techniques, may be leveraged to automate some of these tasks~\citep[e.g.][]{Kerzendorf2017arXiv170505840K}. Astrophysics is full of examples of parameters which may be determined through multiple experimental and observational techniques, and where discrepancies between the resulting values is of particular interest in discussions of the underlying physics. Automating the process of gathering and analysing these measurements would make many avenues of research faster and easier, and open up new possibilities for examining the dissemination of information in the astrophysics community.

To this end we are developing a tool to automatically find, collate, and analyse measurements present in astrophysical literature. The resulting database of measurements would allow for researchers to quickly find an overview of a given parameter, either to find a statistically derived consensus value, or gain an understanding of the distribution of measured values for a given quantity. Such a collection of datapoints -- which would, of course, contain origin publications and potentially other contingent data (experimental technique, for example) -- would also be a excellent starting point for more sophisticated meta-studies and targeted investigations. Additionally, with many papers being submitted to online, open-source repositories, the database may be automatically kept up to date with a minimal amount of manual intervention.

The first step in reaching this goal is an investigation into the available data (textual and catalogue), both in terms of data structure and format, and some examination of the way in which data is presented in scientific writing. Following on from this, models for data extraction must be created, which will highlight important obstacles and future avenues of exploration, which in turn will inform the later implementation of more advanced machine learning techniques. The models we discuss in this paper will primarily be rule-based, and aimed at extracting measurements of named quantities. A ``measurement'' in this context specifically refers to a numerical value with associated uncertainties and units. Concrete examples of measurement reporting from astrophysics publications are given in Examples~1-4 in Table~\ref{tab:examples}.

For the purposes of this work we shall be focusing on finding instances of the Hubble constant in astrophysical texts -- the parameter which describes the expansion rate of the Universe at the current epoch. We have chosen the Hubble constant for two reasons: Firstly, the uniformity of its naming conventions -- both in written English and mathematical syntax -- make it a good test for our explorations into the data. Secondly, the debate over its value -- both historically and in the present \citep{Freedman2017NatAs...1E.169F,Riess+2018ApJ...861..126R,Planck+2014A&A...571A..16P} -- will allow us to check for the presence of expected trends in our results. In future work we shall be extending the method to allow for any named parameter -- even those with linguistically complex names.



In this paper we shall describe our exploration of the astrophysical literature available from the arXiv repository, rules-based models for measurement extraction, and artificial neural network models for measurement classification. We shall begin in Section~\ref{sec:data} with a brief overview of aspects of the data, and move on to Section~\ref{sec:pipeline} to describe our pipeline for producing a unified, easily manipulable set of files. In Section~\ref{sec:measurementextraction} we shall discuss our model for extraction of values of the Hubble constant from arXiv papers, describing the initial model and the improvements required to reduce noise in the output. Using our model we are able to find a strong signal in the data centred around the accepted region for the value of the Hubble constant. Additionally we find structure expected from the current state of the community, notably the two concentrations of results at ${\sim}68$ and ${\sim}73$ km s$^{-1}$Mpc$^{-1}$ seen from 2013 to the present (see Figure~\ref{fig:splithist}). Then in Section~\ref{sec:classifyingarticles} we discuss the training of an artificial neural network classifier for determining if a given paper reports a novel measurement. This is used in conjunction with our extracted values of the Hubble constant to examine the distributions of quoted and novel values in both the time and measurement value axes. Little structure is observed in the time axis, but strong patterns are seen in the value axis (notably a strong peak seen at ${\sim}75$ km s$^{-1}$Mpc$^{-1}$, the accepted region of the true value). Finally, in Section~\ref{sec:conclusion}, we outline future directions for this work, and obstacles which must be overcome in extracting measurements for entities with linguistically complex names.

\begin{figure*}
	\includegraphics[width=0.9\textwidth]{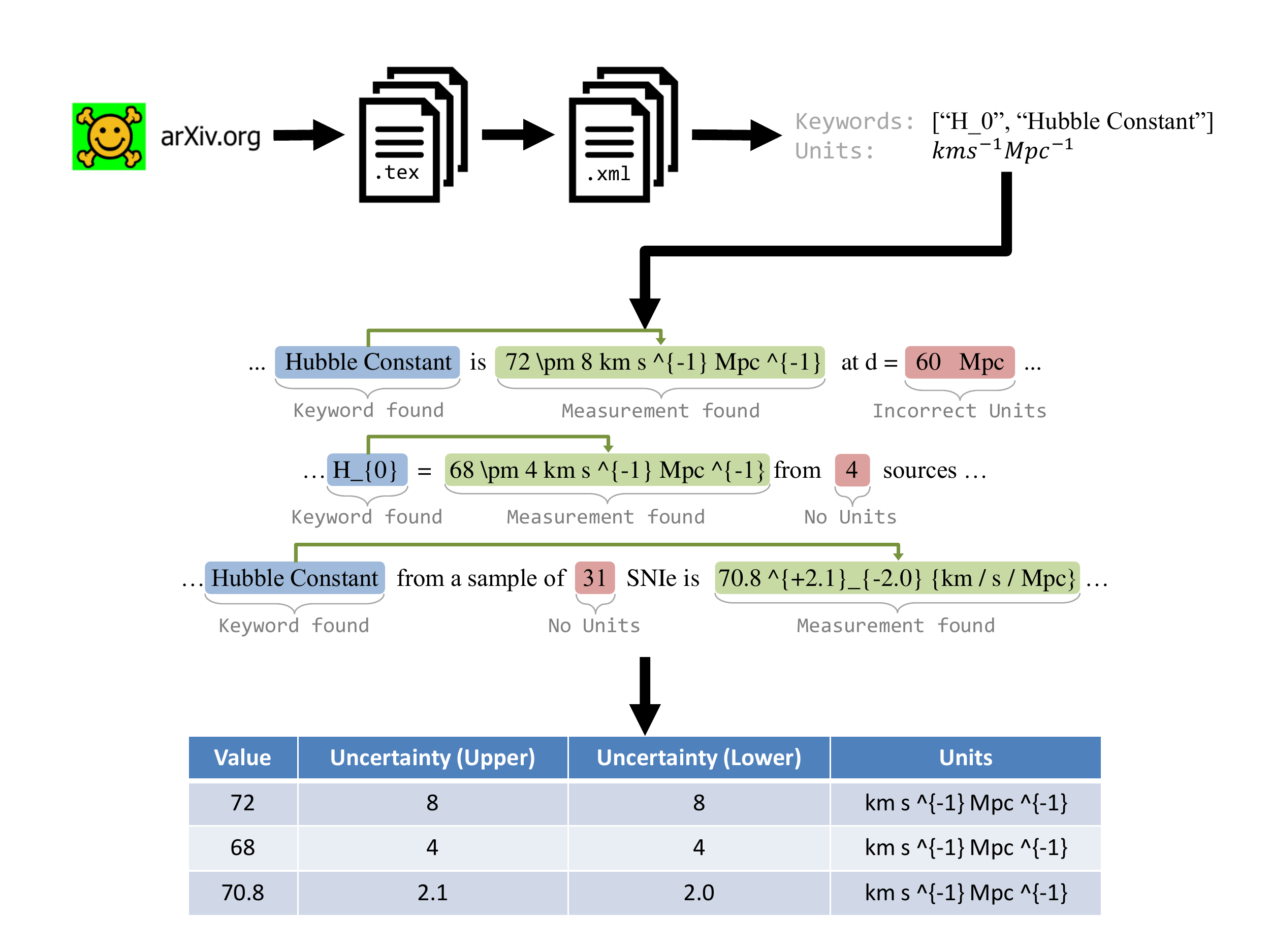}
	\caption{Schematic overview of the project. \LaTeX{} source files are extracted from the arXiv repository, converted into a more practical format (XML), and then spans containing reported measurements of a given entity (in this case the Hubble constant, $H_0$) are identified and processed. The resulting processed data may then be tabulated and analyzed.}
	\label{fig:paperfigure}
\end{figure*}

\begin{table*}
\centering
\caption{Examples of the \TeX{} source for typeset measurement reporting in astrophysical literature, along with the numerical value extracted (and converted to standard units for the Hubble constant, km s$^{-1}$Mpc$^{-1}$) by the models detailed in Section~\ref{sec:measurementextraction}. These examples are all related to attempts to extract the Hubble constant. The arXiv identifier for each source article is provided -- note that all examples originate in article abstracts. The examples have been grouped into the following (in descending order): well formatted examples, well-formed examples which are reporting a different quantity, assumed values of the Hubble constant (i.e. not actual measurements), values related to the Hubble constant (but not measurements), examples where the incorrect number has been identified by the algorithm, and typesetting errors.}
\label{tab:examples}
\resizebox{\textwidth}{!}{%
\begin{tabular}{rrrl}
\hline
Number & arXiv Identifier & Value      & Tokenized \TeX{} Source \\
\hline
\multicolumn{4}{l}{\textit{Well Formatted}} \\
1  & astro-ph/0001156 & 70         & For a flat universe with H \_ \{ 0 \} =70 km s \textasciicircum \{ -1 \} Mpc \textasciicircum \{ -1 \} and q \_ \{ 0 \} = 0.5 \\
2  & astro-ph/0001533 & 74         & H \_ \{ 0 \} = 74 \textasciicircum \{ +18 \} \_ \{ -15 \} ( 95 \% stat. ) \textasciicircum \{ +22 \} \_ \{ -22 \} ( sys. ) km s \textasciicircum \{ -1 \} Mpc \textasciicircum \{ -1 \} \\
3 & astro-ph/0012376 & 72         & consistency with H \_ \{ 0 \} = 72 \textbackslash{}pm 8 km s \textasciicircum \{ -1 \} Mpc \textasciicircum \{ -1 \} \\
4 & astro-ph/0604129 & 70.8       & constraint on the Hubble constant : H \_ \{ 0 \} = 70.8 \textasciicircum \{ +2.1 \} \_ \{ -2.0 \} \textbackslash{}mathrm \{ km / s / Mpc \} \\
\hline
\multicolumn{4}{l}{\textit{Well Formatted - Different Quantity}} \\
5 & 0802.3219         & 13.7       & The result is H \_ \{ 0 \} \textasciicircum \{ -1 \} = 13.7 \textasciicircum \{ +1.8 \} \_ \{ -1.0 \} \textbackslash{}mathrm \{ Gyr \} \\
6 & 1406.7695        & 222        & Hubble parameter data , such as [...] measurement of H ( z ) = 222 \textbackslash{}pm 7 km/sec/Mpc at z = 2.34 \\
7 & astro-ph/0309739 & 0.96       & we find that H \_ \{ 0 \} t \_ \{ 0 \} = 0.96 \textbackslash{}pm 0.04 \\
\hline
\multicolumn{4}{l}{\textit{Assumed Values}} \\
8  & astro-ph/0307223 & 71         & For a cosmological model with H \_ \{ 0 \} = 71 km s \textasciicircum \{ -1 \} Mpc \textasciicircum \{ -1 \} , \textbackslash{}Omega \_ \{ M \} = 0.3 \\
9  & 0705.4505         & 70         & ( when using H \_ \{ 0 \} = 70 km s \textasciicircum \{ -1 \} Mpc \textasciicircum \{ -1 \} ) \\
10 & astro-ph/0112489 & 60         & For all practical purposes H \_ \{ 0 \} = 60 is recommended with a systematic error of \\
11  & astro-ph/0110631 & 70         & adopted Hubble constant of H \_ \{ 0 \} \textbackslash{}simeq 70 \{ km s \textasciicircum \{ -1 \} Mpc \textasciicircum \{ -1 \} \} on the  Hubble diagram \\
\hline
\multicolumn{4}{l}{\textit{Related Values}} \\
12  & astro-ph/0001298 & 65         & the Hubble constant to be H \_ \{ 0 \} \textbackslash{}lesssim 65 \textbackslash{}eta \textasciicircum \{ -1 / 8 \} km/s/Mpc at the two sigma level \\
13 & astro-ph/9909260 & 4          & the derived value of the Hubble constant would increase by 4 km s \textasciicircum \{ -1 \} $\sim$\{ \} \{ Mpc \} \textasciicircum \{ -1 \} \\
14 & astro-ph/9905080 & 3          & an uncertainty of only 3 km s \textasciicircum \{ -1 \} Mpc \textasciicircum \{ -1 \} of the Hubble constant \\
15  & 0705.0354         & 5          & and \textbackslash{}Delta H \_ \{ 0 \} = 5 \% for the Hubble constant \\
16 & astro-ph/0609109 & 25         & to be \textbackslash{}Delta H / H \_ \{ 0 \} \textbackslash{}sim ( 25 \textbackslash{}pm 15 ) \% \\
\hline
\multicolumn{4}{l}{\textit{Incorrect Number Identified}} \\
17 & astro-ph/0112040 & 0.0        & \textbackslash{}Omega \_ \{ \textbackslash{}Lambda \} = 0 , H \_ \{ 0 \} = 50 km s \textasciicircum \{ -1 \} Mpc \textasciicircum \{ -1 \} \\
18  & astro-ph/0110054 & 1          & of \{ T \_ \{ 0 \} \} \textbackslash{}times \{ H \_ \{ 0 \} \}  ;  ( iii ) the Einstein-de Sitter model ( \textbackslash{}Omega \_ \{ 0 \} = 1 , [...] ) \\
19  & astro-ph/0602109 & 0.1        & and z = 0.1 , the value of the estimated H \_ \{ 0 \} is positively biased with \\
20 & astro-ph/0305008 & -1.0       & of the dark energy is w = -1 , then H \_ \{ 0 \} t \_ \{ 0 \} = 0.96 \textbackslash{}pm 0.04 \\
\hline
\multicolumn{4}{l}{\textit{Typesetting Errors}} \\
21 & astro-ph/0210529 & $6.5 \times {10}^{9}$ & H \_ \{ 0 \} = 65 \{ km s \textasciicircum \{ -1 \} mpc \} \textasciicircum \{ -1 \} \\
22 & 0807.0647         & 0.765      & these tests yield H \_ \{ 0 \} = 0.765 \textasciicircum \{ +0.035 \} \_ \{ -0.033 \} km s \textasciicircum \{ -1 \} Mpc \textasciicircum \{ -1 \} \\
\hline
\end{tabular}}
\end{table*}

\section{Data} \label{sec:data}

The arXiv, operated by the Cornell University Library, represents one of the largest open-source repositories of scientific literature available. It has seen considerable uptake in the physical sciences, especially astrophysics, and hence it will be used in this work as a source of text for data extraction and model training purposes.

\begin{figure}
	\includegraphics[width=\columnwidth]{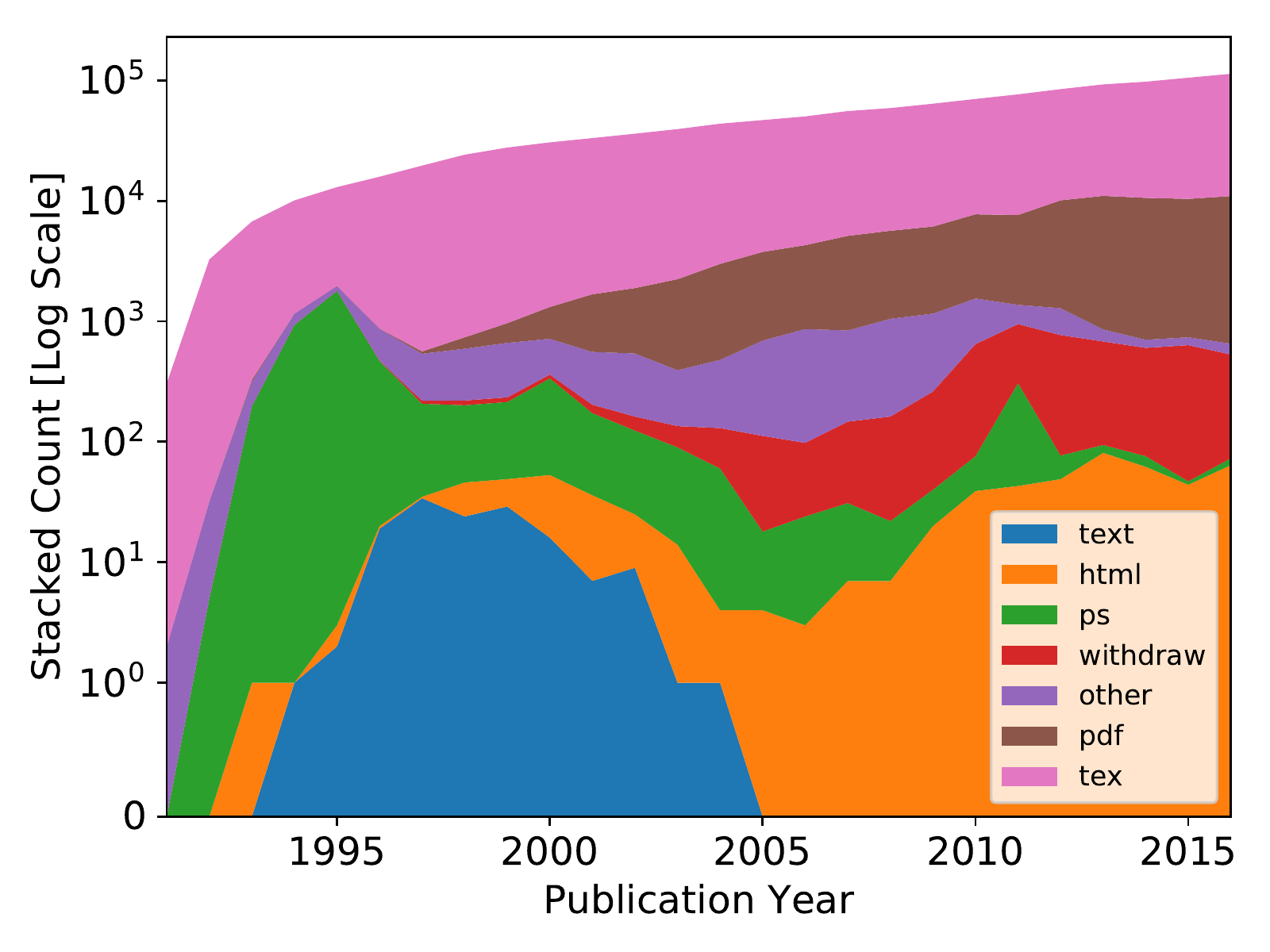}
	\caption{Distribution of arXiv source file groupings (see Table~\ref{tab:sourcetypes}) with time. Group occurrences are plotted using a log-scale. \TeX{}/\LaTeX{} source files dominate the distribution, followed by PDF files.}
	\label{fig:arXivstats}
\end{figure}

The arXiv makes available \LaTeX{} source files for the vast majority of its articles, roughly 91\%, and we shall be focussing on this subset for our preprocessing steps. We investigated the distribution of filetypes (based on file extension) across all the arXiv source files to determine if there was another prevalent filetype which should be accounted for. The source files include all manner of different file extensions, from various \TeX{} and \LaTeX{} extensions (e.g. \texttt{.tex}, \texttt{.TEX}, \texttt{.latex}, \texttt{.ltx}, etc.) to unusual compression formats (e.g. `\texttt{.cry}'), and many others inbetween. Entries without \LaTeX{} source files fall into a number of groupings, such as entirely different source filetypes or withdrawn papers, and a summary of these may be seen in Table~\ref{tab:sourcetypes} and Figure~\ref{fig:arXivstats}. The largest grouping, aside from \TeX{} and \LaTeX{} source files, is for articles available only in PDF format (7.5\%). Due to the complexity of extracting well-formatted textual data from PDFs, we shall exclude such files during preprocessing, operating under the assumption that there is no systematic disparity between the general trend in \LaTeX-submitted papers versus PDF-submitted papers. Verifying this claim is beyond the scope of this paper, and the following results are based on this working assumption.

Our data consists of the source files for all arXiv articles up until September 2017 (the earliest article being from July 1991), corresponding to a total of 1,309,498 articles.
Our preprocessing pipeline (see Section~\ref{sec:pipeline}), which requires that the \LaTeX{} source files be present for the article, yields 208,541 processed astrophysics articles. Of these 195,369 articles (94\%) have an `abstract' section (i.e. the article has made use of the \LaTeX{}-specific `\texttt{\textbackslash abstract}' command), which will be a useful structure in our analysis. The reason for this reduction is that some of the processed articles have \TeX{}-only source files, and therefore cannot include the \LaTeX{} `\texttt{\textbackslash abstract}' command (or many other useful \LaTeX{} structures). Additionally we also find 142,179 articles (68\%) with both an identifiable abstract and conclusion. The conclusions are identified using `\texttt{\textbackslash section}' structures with titles containing either ``conclusion'' or ``summary'' (case insensitive search).

\begin{table*}
\centering
\caption{Distribution of arXiv source file categories, with common file extensions (note that these extensions may employ different capitalisations), descriptions of the categories, and percentage occurrences in arXiv. See Figure~\ref{fig:arXivstats} for a representation of these distributions with time.}
\label{tab:sourcetypes}
\begin{tabular}{llll}
\hline
Category & File Extensions    & Description                                            & Percentage \\
\hline
tex      & \texttt{.tex}, \texttt{.latex}, \texttt{.ltx} & \TeX{} or \LaTeX{} source files present                      & 90.94      \\
pdf      & \texttt{.pdf}               & No source provided, only PDF                           & 7.46       \\
withdraw & N/A                & Source contains only filenames containing ``\texttt{withdraw}'' & 0.39       \\
ps       & \texttt{.ps}                & All files in PostScript format                         & 0.38       \\
html     & \texttt{.html}              & All files in HTML format                               & 0.05       \\
text     & N/A                & Source contains only file(s) named ``\texttt{text}''            & 0.01       \\
other    & N/A                & Unusual source directory                               & 0.76       \\
\hline
\end{tabular}
\end{table*}

In addition, we have utilised the dataset compiled by \citet{Croft+Dailey2011arXiv1112.3108C} as validation data and a source of example literature in this work. The dataset consists of 638 compiled values of 8 cosmological parameters from 468 papers. Of these, 214 papers (46\%) are successfully processed by the pipeline described in Section~\ref{sec:pipeline}. More specifically, 124 of the 638 measurements in this dataset (19\%) are Hubble constant measurements, originating from 122 of the 468 papers (26\%). Of these 122, 80 papers (65\%) are successfully processed by our pipeline. The low efficiency for the conversion of these papers is due to the dataset being biased towards older publications, which either do not have \LaTeX{} source files (e.g. source is in PostScript format), or otherwise are unusually formatted due to lack of standardisation. These papers in this dataset are used as a starting point for examining occurrences of astrophysical measurements in literature, and also as a gold-standard dataset (albeit single-class) for validation of classifiers in Section~\ref{sec:classifier}.

\section{Pipeline} \label{sec:pipeline}


\LaTeX{} files are not ideal for natural language processing tasks, as they contain a large amount of information which is of use only in type-setting contexts. However, information relating to document structure is of great use when manipulating and analysing the text contained in the article -- for instance, the ability to distinguish sections, easily identify article abstracts, and so on. As such, we require a document format into which the \LaTeX{} source files can be converted which will retain the structural information we desire, but will facilitate ease of access in computational settings. To this end we employ LaTeXML\footnote{LaTeXML homepage: \url{http://dlmf.nist.gov/LaTeXML/}}, a program which converts \LaTeX{} files (including style and class files, thus accounting for custom commands and macros) into XML format. The hierarchical structure of XML is well suited to representing the structure of scientific literature, where articles contain sections which themselves contain subsections and then paragraphs and so on, and the high availability of XML libraries in all major programming languages make this document format a desirable choice for our purposes.



File extensions are used to find the required documents from the arXiv source directories (discounting figures and other unnecessary files). As mentioned earlier, this leads to some issues with the large variety of extensions employed by writers, with Table~\ref{tab:sourcetypes} indicating the assumptions that have been made here when identifying \LaTeX{} source files by extension. The preprocessing pipeline then processes each article's source files in the following steps:

\begin{enumerate}
\item Article category tags are found from the arXiv metadata, and articles without the astrophysics tag (``\texttt{astro-ph}'') are discounted.
\item Article source files which match known \TeX{}/\LaTeX{} file extensions (e.g. \texttt{.tex}, \texttt{.cls}, \texttt{.sty}, \texttt{.bib}) are identified.
\item If more than one \TeX{} file is present, each file is scored to determine the main source file. This step is more complex than expected, as it transpires that many source directories contain more than one file with a ``\texttt{\textbackslash begin\{document\}}'' expression. Presence of the ``abstract'' keyword and the article title (taken from the arXiv metadata) are used in this scoring. Approximate string matching is used to find the article title, due to the discrepancies which may be found between titles stored in the metadata, and that which appears in the source text, often due to the presence/absence of mathematical type-setting commands.
\item The highest scoring file is processed using LaTeXML.
\item The text stored in the XML tree is tokenised and sentence split, such that all words and punctuation tokens are separated with whitespace, and each line contains a single sentence (and sentences are not split between multiple lines). This stage facilitates use of the data in a natural language processing context.
\end{enumerate}


When run on the arXiv source dataset this process yields 208,541 astrophysics articles in XML format, with a total of 12,868 failures due to decoding or LaTeXML errors, giving a success rate of 94\%. This is considered sufficient coverage for our purposes.

\section{Measurement Extraction} \label{sec:measurementextraction}

We now wish to produce an algorithm for extracting measurements from text. There exist many machine learning techniques in the natural language processing domain for this class of problem (e.g. named-entity extraction, question-answering, etc.) that we may apply in this scenario, however we shall begin by producing a baseline model: a simpler model which trades effectiveness for legibility, based on techniques which may be easily reasoned about. The output of this model may then itself be used as a baseline when experimenting with more complex models -- this, indeed, shall be the subject of future work (see Section~\ref{sec:conclusion}) -- and hence will be a good test of these models' effectiveness.

We shall begin with a method of measurement extraction based on a simple keyword search. Given our processed arXiv articles it is a simple task to search for a specified keyword in the document, and instances of numerical values. We then make our primary assumption: that the closest numerical value to a keyword instance is a measurement of the entity to which the keyword refers. This is a strong assumption, but shall be seen to produce useful results. The next assumption we shall make is that numerical values and the names of the entities to which they refer are found in the same sentence -- i.e. there is no multi-sentence inferencing required. Examination of real-world scientific literature shows that neither of these assumptions holds in all cases, but as a general trend they are a good starting point for our model.

Here we shall focus on extracting measurements of the Hubble constant from the arXiv astrophysical literature dataset. The Hubble constant is a good candidate for this type of keyword search as it has a small number of recognisable identifiers which differ little between authors. Notably, we have the following:
\begin{itemize}
\item Hubble constant
\item Hubble parameter
\item $H_0$: written `H\_0', `H\_\{0\}', `H\_\textbackslash circ', or `H\_\{\textbackslash circ\}'
\end{itemize}
with optional capitalisation of the second word in the above phrases. These may easily be encoded by hand if one has some knowledge of the typesetting conventions for the common mathematical symbol.

We shall also be focusing primarily on measurements extracted from article abstracts. Our reasoning for this is as follows: at a pragmatic level, experimentation shows that paper abstracts include far fewer extraneous or arbitrary numbers than the article bodies. These numbers may include: year dates from citations, section/equation reference numbers, secondary calculated values, assumed values, and so on. Limiting the search to article abstracts greatly reduces noise in the output, whilst preserving values of interest. This is motivated with the assumption that any paper whose main subject is the measurement of some physical quantity will give a summary of said measurement in its abstract. Similar approaches have been taken in data extraction work in the bio-medical field \citep{pmid12967967,Usami:2011:AAH:2002902.2002912}. Based on observation of scientific literature we would expect these summaries to be of the form ``we find \textit{name} to be \textit{value}$\pm$\textit{uncertainty}'', or ``\textit{symbol} = \textit{value}$\pm$\textit{uncertainty}'', or similar. Note that there are, of course, many variations of these patterns, and the models discussed below are designed to be as robust to them as possible.

For clarity, we shall list the above assumptions here:
\begin{enumerate}
\item Closest numerical value to a keyword instance is a measurement of the entity to which the keyword refers.
\item Numerical values and associated entity names appear in the same sentence.
\item Values of interest appear in the article abstract.
\end{enumerate}

\subsection{Initial Model} \label{sec:initialmodel}

It transpires that the naive application of our assumption of taking the closest number to a keyword produces a large amount of noise. There are simply too large a variety of ways a simple series of digits (and possibly a decimal point) can occur in a sentence -- especially in scientific text, which contains many numerical identifiers (e.g. ``NGC1277'' for a galaxy, or ``0703.00001'' for an arXiv identifier), and mathematical expressions. For example, consider the following strings: ``H\_\{0\}', ``H\_\{z=1.5\}'', ``a=b-1'', ``a=1-b'', and so on. Patterns such as these are common in scientific writing. We may solve the first two by simply assuming that all numbers enclosed in braces (``\{ \}'') are related to \LaTeX{} math expressions and not numerical values in their own right. The latter two present more of an issue, however, as it is not evident that a simple rule may be constructed to remove them which would not also interfere with finding actual measurements.

However, there do exist some simple patterns which we may account for. Any numerical string returned by the initial search for numbers in the text which overlaps in the sentence with one of the following patterns is rejected as a possible measurement:
\begin{itemize}
\item Year date, expressed as a series of 4 digits in parentheses, where the resulting value lies in the range 1400-2100, e.g. ``(1990)''
\item Year date followed by proper noun (capitalised word), e.g. ``2013 Planck''
\item Identifier (any digits preceded by an uppercase string), e.g. ``NGC1277''
\item ArXiv identifier, e.g. `astro-ph0101001' or `0703.00001'
\end{itemize}
These filters greatly reduce noise in certain numerical ranges (notably 1980-2020, the standard range for references in modern scientific literature), and generally reduce the number of outliers.

\begin{figure*}
    \subfloat[Initial Model\label{fig:initialModel}]{
        \includegraphics[width=0.45\textwidth]{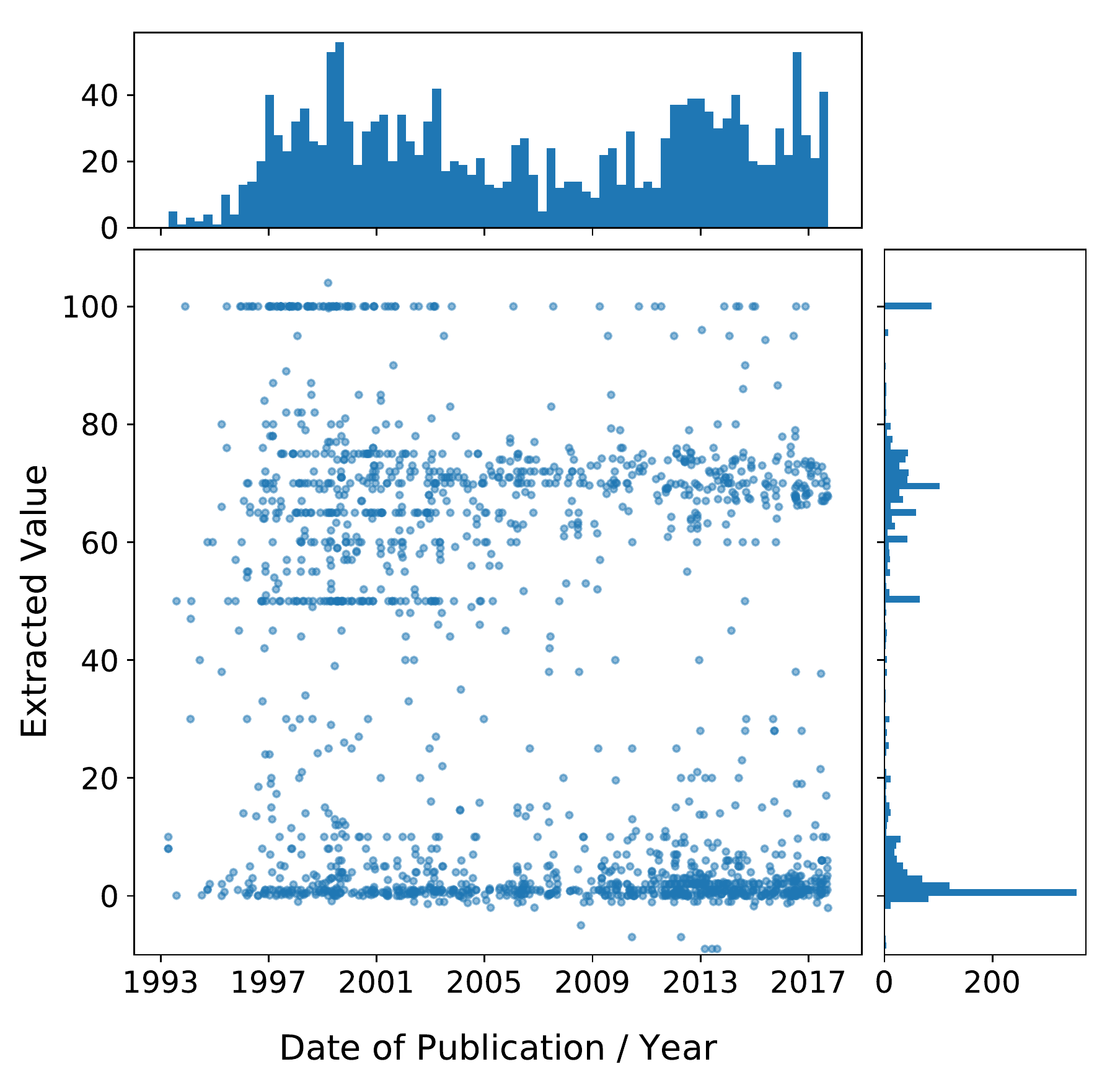}
    }
    \hfill
    \subfloat[Improved Model\label{fig:improvedModel}]{
        \includegraphics[width=0.45\textwidth]{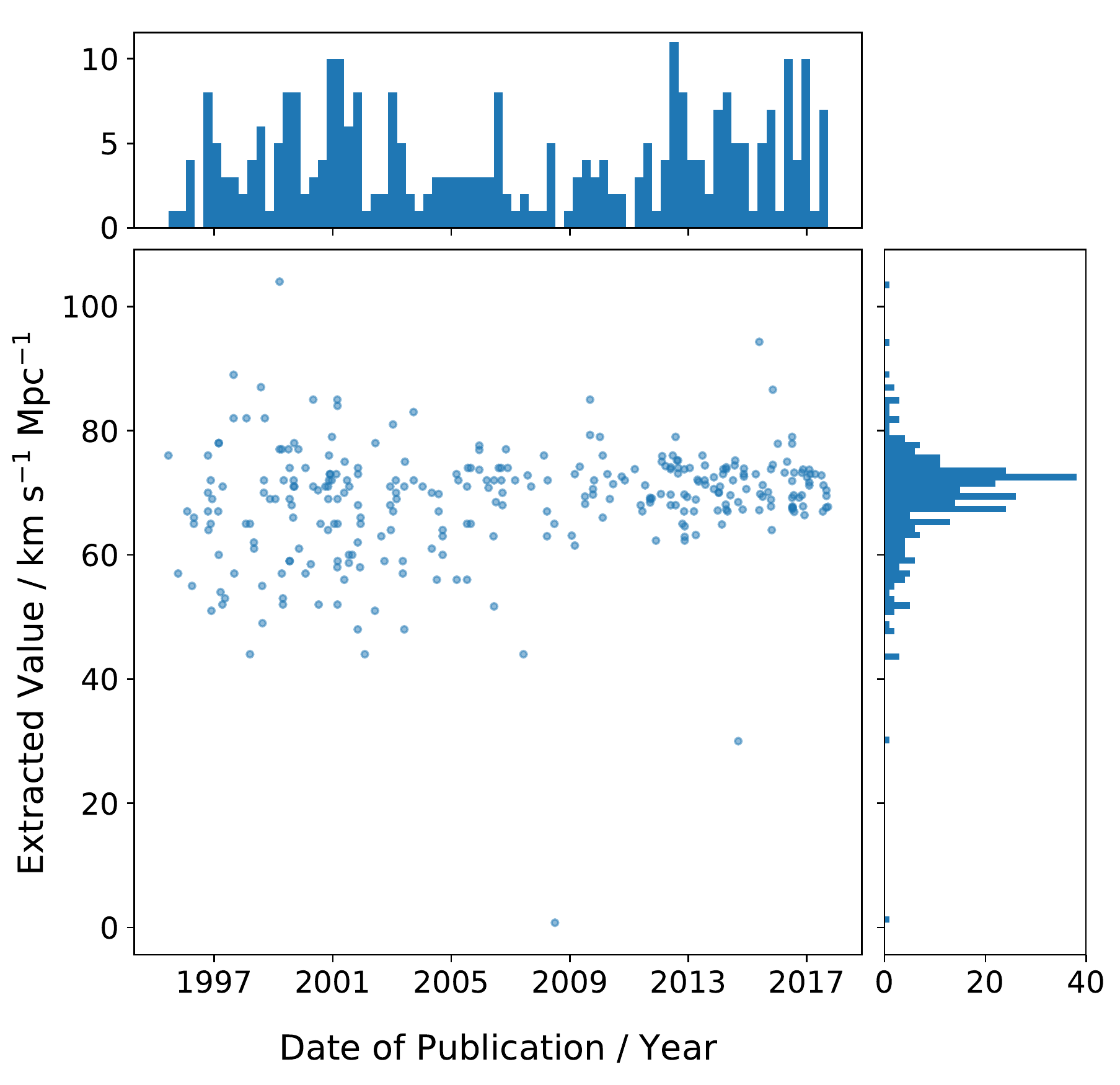}
    }
    \caption{Outputs of models at different stages of development. Time- and value-domain histograms are also shown. Plot~\protect\subref{fig:initialModel} shows the output of the initial model. This plot shows all numbers matched to keyword instances in available arXiv astrophysics papers, using the approach described in Section~\ref{sec:initialmodel}. The groupings at 0, 50, and 100 in the measurement axis are particularly notable, with the grouping at 0 primarily consisting of noise. Plot~\protect\subref{fig:improvedModel} shows the output of the improved model. This plot shows all measurements (numerical values reported with an uncertainty and the correct dimensions) matched to keyword instances in available arXiv astrophysics papers, using the approach described in Section~\ref{sec:improvedmodel}. Here we may note the absence of the assumed values at 50 and 100 km s$^{-1}$Mpc$^{-1}$, and the noise around 0 on the measurement axis.}\label{fig:modeloutputs}
\end{figure*}

Using the above written forms of the Hubble constant and the practical additions to the search method, we shall perform our search on the available astrophysical literature. This returns 1730 values from 1324 paper abstracts. The results are shown in Figure~\ref{fig:initialModel}. Note that, for the sake of readability, 5\% of the returned data lies outside the range of the figure (corresponding to 93 values).

The most striking issue with the plot is the large cluster of values around 0. These are mostly caused by the search algorithm being overly-generous when searching for numerical values, or by a failure of one of our earlier assumptions. For instance, we may find a keyword in a sentence which does not actually report a measurement of the keyword, but which does contain other numerical data, such as Example~19 in Table~\ref{tab:examples}. Or where the arrangement of characters in the sentence causes the wrong number to be interpreted as the ``closest'' (where grammatically the reader would understand the relationship, but our simple algorithm cannot), such as Example~17 in Table~\ref{tab:examples}. We may also find a different use of one of our keywords, such as in a compound quantity involving a mathematical keyword -- for example, ``H \_ \{ 0 \} t \_ \{ 0 \}'' in Example~7 in Table~\ref{tab:examples}. It should be noted that these issues also lead to noise in other numerical ranges, but the nature of scientific literature (or, at least, astrophysical literature) seems to lead to values around ${\sim}0$ appearing with great frequency in text. Many of these are found to be literary devices (e.g. section numbers), or digits in equations (e.g. $x=1-y$).

We may also note the strong lines present at 50 and 100 km s$^{-1}$Mpc$^{-1}$. These are common assumed values for the Hubble constant. Their presence (and the presence of other such assumed values) is discussed in Section~\ref{sec:results}.

\subsection{Improved Model} \label{sec:improvedmodel}

The largest issue with the above form of the search is in the way numerical values are identified (i.e. the characters in the string which correspond to numerical values). Simply filtering out numbers which appear inside mathematical symbols and common non-measurement patterns is insufficient. The next step shall be to produce a more sophisticated regular expression for identifying numerical values in text -- specifically numerical values which are a part of a measurement. A common signifier of a scientific measurement is the presence of an uncertainty, and we shall take advantage of this to filter out non-measurement numerics.

First we must consider the standard patterns used to report such measurements. Examination of the literature yields the following common patterns:
\begin{itemize}
\item Plus-minus symbol: $1.0\pm0.5$
\item Upper and lower bounds: $1.0^{+0.1}_{-0.2}$
\item Named uncertainties: $1.0^{+0.1}_{-0.2}$ (random) $\pm0.3$ (statistical)
\end{itemize}
and combinations and repetitions thereof. There are, of course, other more complex patterns which occur frequently, but these represent the most common and easily codifiable, and hence shall be our starting point. These may be encoded into a regular expression which is used to identify measurement patterns in the text, which may then be matched to the nearest keyword instance, as before. We may now specify that a numerical value must be followed by an uncertainty to be considered a `measurement'.

Further to this we may wish to specify the dimensions of the measurement we are searching for. Once again we may construct a regular expression, now to search for units following a number (potentially with included uncertainties). This may be done by simply assuming all \LaTeX{} math symbols and tokens consisting of less than 3 characters following a number are part of its units. A simple context-free grammar may then be used to parse the string returned by the regular expression -- as our regular expression is becoming rather cumbersome at this point. This final parsing is also used to remove any extraneous characters from the end of the string, and convert the measurement into a standardised format which may be more easily processed. The use of the context-free grammar and this standardisation allows for a variety of mathematical syntax to be accepted in the units string -- for example, ``km s$^{-1}$ Mpc$^{-1}$'' and ``km/s/Mpc'' are equivalent in our search, and both would be equivalent to ``s$^{-1}$'' (given appropriate numerical conversions).

We may now specify that for a number to be considered a ``measurement'' it must possess both an uncertainty, and a given dimensionality. Running this search for the Hubble constant, and specifying units of km s$^{-1}$Mpc$^{-1}$, we find 295 measurements from 225 paper abstracts. The results are shown in Figure~\ref{fig:improvedModel}. Note, only 1 value now lies outside the plotted region, which corresponds to Example 6 in Table~\ref{tab:examples}, as discussed below.

To summarize, we are now using the following assumptions:
\begin{enumerate}
\item A numerical value cannot be a measurement if it is contained within a pattern for a date or identifier (see Section~\ref{sec:initialmodel} for concrete rules).
\item A numerical value is a potential measurement if it appears with an uncertainty and the expected dimensions.
\item The closest such numerical value to a keyword instance is a measurement of the entity to which the keyword refers.
\item Numerical values and associated entity names appear in the same sentence.
\item Values of interest appear in the article abstract.
\end{enumerate}

Our previous issues have now been mostly tackled successfully, but a greater problem is now presented by author error.
For instance in Example~22 the author has confused their results for $H_0$ and little $h$ (where $h=H_{0}/100$ [km s$^{-1}$Mpc$^{-1}$]), thus leading to an incorrect statement of their measurement - it should be noted that the result is correctly reported elsewhere in the paper. Examination of the outliers present in this plot confirms that each one is either an author syntax error, or a genuine report of an unusual value. It should be noted that these unusual values are often reported alongside more expected values in the same section -- for example where different techniques, or inclusion of some additional physics to a model, produce a significantly different result.

We may also note the absence of the 50 and 100 km s$^{-1}$Mpc$^{-1}$ lines. This is to be expected, as these values are rough estimates, and hence are generally not reported with any kind of uncertainty. They are, however, usually reported with the correct units -- and these lines would indeed reappear if we required only the presence of the correct units, but not an uncertainty. An example of this may be seen in Figure~\ref{fig:prettyPlot} later in this work.

\section{Classifying New Measurements} \label{sec:classifyingarticles}

In addition to finding and extracting instances of reported measurements in text we also wish to differentiate between quoted values (from some previous work) and newly reported values (i.e. the results of original work presented in the paper). Both are of interest for different purposes: we may wish to measure the popularity of certain values, as well as find and plot the progression of new values. To begin we shall simply attempt to classify papers by whether or not each paper reports any new measurements. Papers which do report new measurements shall be considered positive samples, and papers which do not (but which may still be quoting pre-existing values) shall be considered negative samples.

For this classification task we shall be utilising machine learning algorithms (specifically artificial neural networks) as opposed to the rules-based approach we employed in our measurement extraction above. This is due to two primary reasons: firstly, producing rules to distinguish positive and negative samples is a very difficult task, as the linguistic and structural cues are complex and hard to codify (in part because they often extend over multiple parts of the text). It is, however, possible to construct rules which may select positive samples with high precision and low recall (i.e. many false-negatives), which may be used to construct a training dataset, as discussed below. Using such a training dataset we can attempt to generalise from our initial assumptions, and uncover patterns we could not easily have codified. Secondly, many machine learning algorithms (e.g. neural networks) may be used to produce probabilistic outputs, which is useful in analysis and in prioritising data samples for investigation. As an example, the latter will be useful in identifying promising samples for annotation in future work.

\subsection{Silver Data} \label{sec:silverdata}

Before we train any type of classifier we must first produce a training dataset from our arXiv XML data. Here we shall produce a silver-standard dataset for training purposes -- a "silver" dataset being one where the labels are assigned based on heuristics, as opposed to a "gold-standard" dataset where the labels are assigned manually by a human. It should be noted that the \citet{Croft+Dailey2011arXiv1112.3108C} dataset mentioned earlier is available as a small gold-standard dataset (with some selection bias) for validation purposes. This approach of using heuristics on a large, unlabelled dataset, coupled with a smaller gold dataset, is an effective substitute for large training datasets when training initial/baseline models in machine learning contexts \citep{Mintz:2009:DSR:1690219.1690287}.

For this task we are primarily concerned that our silver dataset have a high precision, which may be attained at the expense of recall. In practice this means we require a set of hand-crafted rules which can positively identify articles which report a new measurement with a high degree of precision (i.e. with the minimal number of false-positives), but where the number of false-negatives (articles which do report a new measurement but are reported as negative samples) may be high. Such a set of rules would provide the positive training samples for our classifier. To find the negative samples we make the assumption that the large majority of papers are not reporting a new measurement value (negative samples), and hence a random sample of the negative articles from the silver data (those deemed by our hand-crafted rules as being negative) should primarily consist of true-negative articles. In this manner we may construct a balanced training data set.

The question now is how to construct the rules which will produce our silver-standard data: As discussed in Section~\ref{sec:measurementextraction}, it is decided that the classifier shall use article abstracts as input data. Hence we must look to other sections of the document to base our rules: after the abstract, the next logical locations would be the title and conclusion. Experimentation with different setups and rules leads to the conclusion that the optimal strategy is to use a combination of these two. The procedure for identifying positive samples is as follows:
\begin{enumerate}
\item The presence of recognisable abstract and conclusion passages is verified (otherwise the document is rejected and shall not be considered for inclusion in the training data).
\item The article title is checked for the presence of at least one of the following words:
\begin{itemize}
\item measurement
\item measuring
\item determination
\item determining
\item estimation
\item value
\item parameter
\item constraint
\end{itemize}
\item The measurement pattern described in Section~\ref{sec:initialmodel} is used to search the conclusion text, and a list of any measurements present is found.
\item Each measurement is checked for the presence of an uncertainty.
\end{enumerate}
If all of the above steps produce a result (i.e. we find one of the listed keywords in the article title, and a measurement with an uncertainty is present in the conclusion), then the article is assumed to be reporting a new measurement and is added to the list of positive samples to be used in training. It should be noted that we are not limiting ourselves to articles reporting a value of the Hubble constant -- any measured value is considered. This method has the advantage of relative simplicity, as it does not rely on phrases or more complex linguistic patterns, but only on word inclusion for the title and pattern matching of \LaTeX{} mathematical notation (a much more formalised and hence codifiable series of tokens) for the conclusion.

However, this simplicity is only advantageous if it works. Manual classification of a sample of the resulting silver data is conducted to test the precision of the model: 200 articles evenly distributed between positive and negative (according to the silver-algorithm) are classified based on the article abstract (note: without the article title) by one of the authors. The resulting manual classifications give a total accuracy of 82\% for the silver algorithm over the 200 samples, corresponding to a precision for the 100 silver-positive samples of 88\%. This is considered sufficient for our purposes, and hence the silver dataset shall be used as training data for our ``new measurement'' classifier.

In total, 1612 positive samples are identified using the above rules.

\subsection{Classifier} \label{sec:classifier}

We shall use an artificial neural network (ANN) classifier to classify articles by whether or not they report a new measurement. We have chosen to use ANNs as they are a standard algorithm in modern machine learning, and shallow networks of the type we shall use here are well studied and understood.

For the input to the model we shall use the article abstracts. Paper abstracts are used for the reasons discussed earlier in Section~\ref{sec:initialmodel}, as they represent a summary of the article contents. This is necessary as using the entire paper leads to the training signal being too weak and the model not learning effectively.

The abstract texts shall be converted into document matrices using a word2vec model specially trained on the entire arXiv astrophysics corpus. Word2Vec \citep{DBLP:journals/corr/abs-1301-3781} is a group of models which allow us to pre-train vector representations of words informed by the entire corpus, which leads to greater generalisation of resulting models trained using these embeddings. This is done by attempting to assign each word in a vocabulary to a vector such that ``similar'' words are close together in the vector space. Words are considered to be ``similar'' if they are found in similar contexts -- i.e. they are often surrounded in a sentence by the same words. In practice we may consider that two words are similar if they are interchangeable in a sentence. For example, we might expect the words ``galaxy'' and ``star'' to both appear in sentences containing the words ``telescope'' and ``observed'' -- in the sentence, ``I observed the the galaxy through the telescope'', we could replace ``galaxy'' with ``star'' and the sentence would still be reasonable (i.e. has a high probability of appearing in our corpus). However, if we replace the word ``galaxy'' with the word ``potato'', the sentence becomes very unlikely. And so our word embeddings for ``galaxy'' and ``star'' are similar, but both are different to our embedding for ``potato''. Using these embeddings, we may now define distance metrics to compare the similarity between word pairs (cosine distance is commonly used for this purpose), and other such mathematical operations.

Hence, using the trained astrophysics word2vec model, the document matrices for the article abstracts are created by concatenating the resulting word-vectors into a single matrix. In our trained word2vec model the word-vectors have dimensionality $d=100$.

The structure of the classifier network is as follows:
\begin{itemize}
\item For an article with an abstract with word-count $n$, a document matrix $D$, of dimensionality $d \times n$, is constructed.
\item The document matrix is multiplied with a (trainable) projection matrix, $P$, of dimensionality $d \times d$, producing the projected document matrix $\tilde{D} = P\times D$.
\item The minimum, maximum, and mean are taken along the rows of $\tilde{D}$ and concatenated to produce a single vector, $x$, of dimensionality $3d$.
\item The vector $x$ is now fed into single dense layer with a single output, as in: $y = \mathbf{w} \cdot \mathbf{x} + b$
\item The output of the dense layer is passed to a sigmoid function to produce the final output of the classifier.
\end{itemize}

Using this setup and the silver dataset described in Section~\ref{sec:silverdata} we may now train our classifier. The dataset is divided into training and testing datasets, with a 90/10\% split, resulting in 1394 each of positive and negative samples for the training set, and 154 for the testing set (these numbers are determined by the number of positive samples found by our rules from Section~\ref{sec:silverdata}). This does not include the validation data points from \citet{Croft+Dailey2011arXiv1112.3108C}. We use the ADAM optimizer, a standard ANN optimizer, along with mini-batching (32 samples per batch), for 100 epochs of training. For each epoch the negative training data is resampled from the available articles (as discussed in Section~\ref{sec:silverdata}), maintaining class balance with the positive training data, resulting in a better coverage of the data over the course of training and exposing the model to a richer set of negative samples. The training was conducted with cross-entropy loss with L2 regularisation, another standard technique in current machine learning. This ANN was implemented using the Flux machine learning library \citep{innes:2018} for the Julia programming language \citep{Bezanson2017}.

It should be noted that longer training runs have been conducted, but the model accuracy and loss are roughly stable from 100 epochs out to 500 epochs. From this we see a final test accuracy of ${\sim}78$\% (true for both the final model of 100 and 500 epoch training runs). Here we are using a prediction threshold of 0.5 for the model. This may not be optimal, given the class-balanced training data (albeit with increased relative coverage of negative samples). However optimisation of this threshold is beyond the scope of this work, as the implied trade-off of recall and precision is application-dependent. For our purposes, we achieve reasonable accuracy with the standard 0.5 cutoff.

To evaluate the performance of our classifier we use the \citet{Croft+Dailey2011arXiv1112.3108C} dataset and the 200 samples manually classified as validation data for the silver-algorithm (see Section~\ref{sec:silverdata}). It should be noted that the \citet{Croft+Dailey2011arXiv1112.3108C} dataset is slightly biased, and single-class, given its focus on a specific domain (i.e. cosmology). The manually classified data contains 113 positive and 87 negative ground-truth samples.
Both of these datasets were excluded from the training data provided to the classifier. We find that the model recovers 87\% of the \citet{Croft+Dailey2011arXiv1112.3108C} dataset publications, compared to 30\% for the silver-algorithm (adjusted for papers available after preprocessing). The model also achieves an accuracy of 88\% over the 200 manually classified samples -- corresponding to a 92\% precision and 86\% recall (for comparison, the silver-algorithm had an 88\% overall accuracy, with 88\% precision and 78\% recall). This indicates that the model may generalise beyond the silver-standard training data (which is a very limited approach, recovering only 1612 samples from the entire arXiv corpus), and may distinguish both positive and negative samples to a reasonable degree of accuracy.

\section{Final Results} \label{sec:results}

\begin{figure*}
	\includegraphics[width=\textwidth]{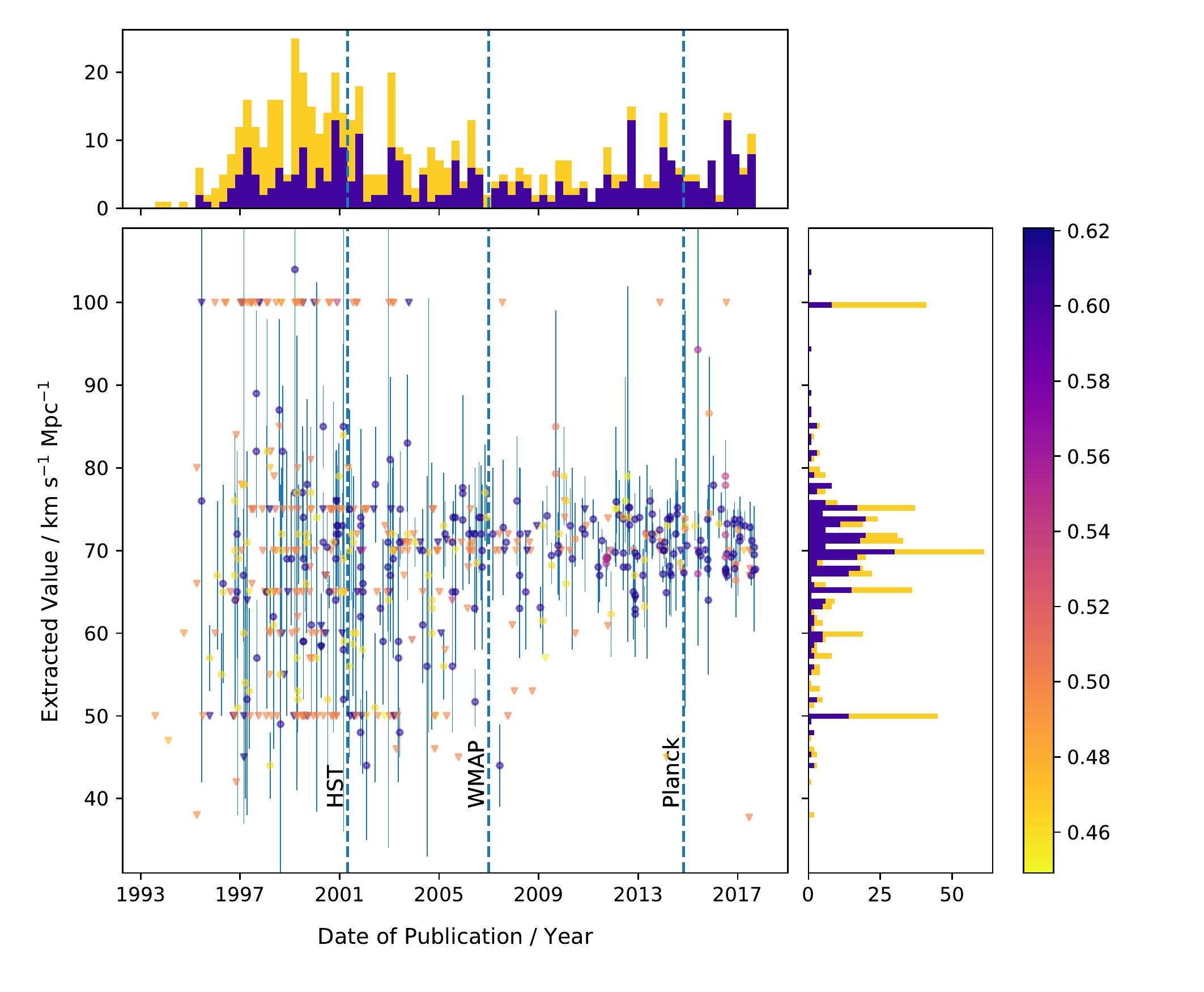}
	\caption{Plot combining output from the improved measurement extraction algorithm and the ``new measurement'' classifier, showing all extracted numbers with the correct dimensionality (km s$^{-1}$Mpc$^{-1}$) from arXiv astrophysical paper abstracts. Datapoint symbols are used to indicate presence of an uncertainty in the reported measurement (circle if present, triangle if not present), with the available uncertainties displayed using error bars. Symbol colour indicates the output of the new-measurement classifier, interpreted as a probability of the measurement originating in a paper reporting a novel value -- colourbar to the right indicates probability value. The stacked histograms indicate distribution in the time- and value-domains (top- and right-hand panels, respectively), with the blue histogram corresponding to measurements whose probability of being a novel measurement is greater than 0.5, and the yellow histogram for the remainder (likely quoted values). The vertical lines correspond to the year of the publication of the HST key project \citep{Freedman+2001ApJ...553...47F}, 3-yr Wilkinson Microwave Anisotropy Probe (WMAP) results \citep{Spergel+2007ApJS..170..377S} and the 2013 Planck results \citep{Planck+2014A&A...571A..16P}.}
	\label{fig:prettyPlot}
\end{figure*}

\begin{figure}
	\includegraphics[width=\columnwidth]{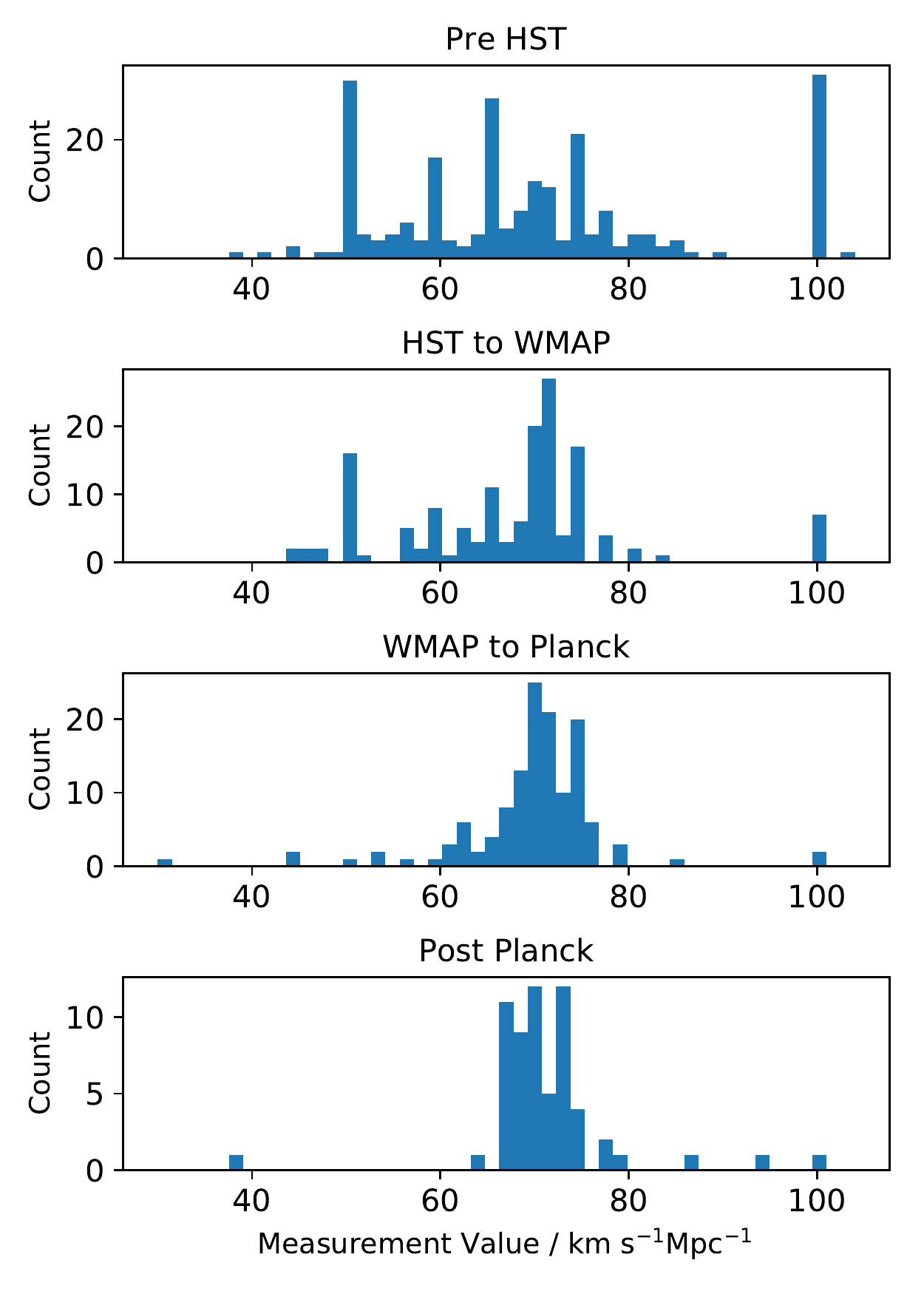}
	\caption{Histograms of the values from Figure~\ref{fig:prettyPlot} between the publication dates of key papers \citep[][ ``HST'', ``WMAP'', and ``Planck'' on the plot, respectively]{Freedman+2001ApJ...553...47F,Spergel+2007ApJS..170..377S,Planck+2014A&A...571A..16P}. We may note the decrease in the spread of reported values over time, along with the decrease in use of the 50 and 100 km s$^{-1}$Mpc$^{-1}$ assumed values, and the eventual disagreement in the value of the Hubble constant post-Planck, as demonstrated by the two peaks at ${\sim}68$ and ${\sim}73$ km s$^{-1}$Mpc$^{-1}$ (the peak at 70 is due to the most common assumed value during this period).}
	\label{fig:splithist}
\end{figure}

We may now combine the results of our keyword-based search with the output of our new-measurement classifier to examine the development of reported values of the Hubble constant in the arXiv literature. To this end we plot found values of the Hubble constant with correct dimensions (km s$^{-1}$Mpc$^{-1}$), both with and without reported uncertainties, which appear in article abstracts, for all viable papers (i.e. the 195,369 papers which have a recognisable abstract section), and the result is shown in Fig~\ref{fig:prettyPlot}. The vertical lines in the figure correspond to the dates of three key publications in the field, to give context to the timeline: the HST key project \citep{Freedman+2001ApJ...553...47F}, the 3-yr Wilkinson Microwave Anisotropy Probe (WMAP) observations \citep{Spergel+2007ApJS..170..377S}, and the Planck 2013 results \citep{Planck+2014A&A...571A..16P}. It should be noted that there are additional outliers outside the bounds of this plot, corresponding to 1.6\% of the available data (9 samples). Of these, 2 are author error, 1 is a historical value (``${\sim}250$ km s$^{-1}$Mpc$^{-1}$''), 1 is a value of $H(z)$ at a different redshift, 3 are uncertainties reported separately to their measurement (with units given), and 1 is a reported change in the value of the Hubble Constant were a different assumption made in the model \citep[Example~13 in Table~\ref{tab:examples}]{2000ApJ...529..786M}, and 1 is a reported difference between local and global measurements \citep{doi:10.1093/mnras/stx1967}. In total we find 573 values from 477 article abstracts. The same data may be seen in Figure~\ref{fig:splithist}, divided into the periods before, after, and between the key publications mentioned above. A few notable features of these plots are outlined below.

\begin{figure*}
    \subfloat[Entire Dataset\label{fig:allsigma}]{
        \includegraphics[width=0.3\textwidth]{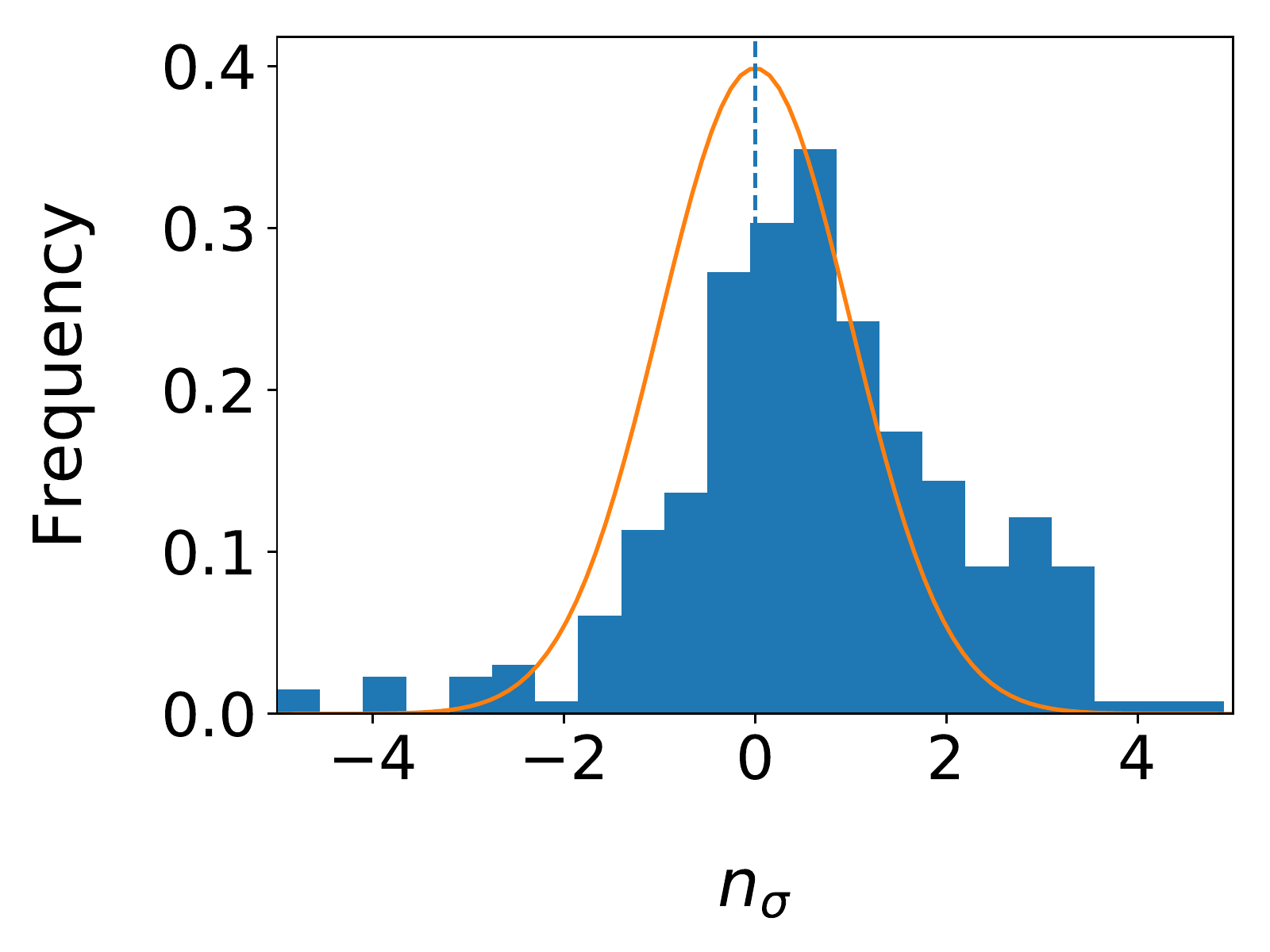}
    }
    \hfill
    \subfloat[Before Planck 2013\label{fig:preplancksigma}]{
        \includegraphics[width=0.3\textwidth]{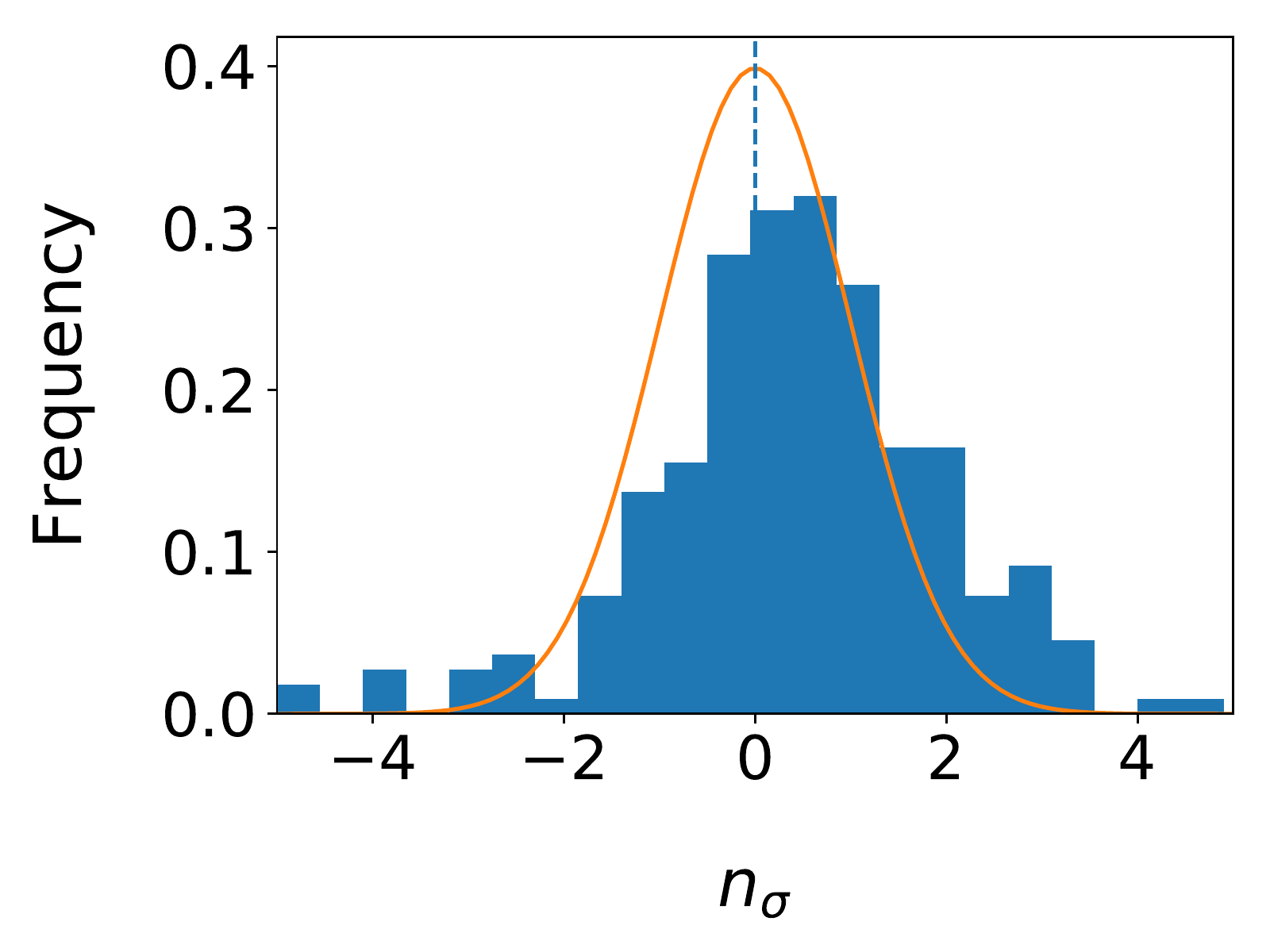}
    }
    \hfill
    \subfloat[After Planck 2013\label{fig:postplancksigma}]{
        \includegraphics[width=0.3\textwidth]{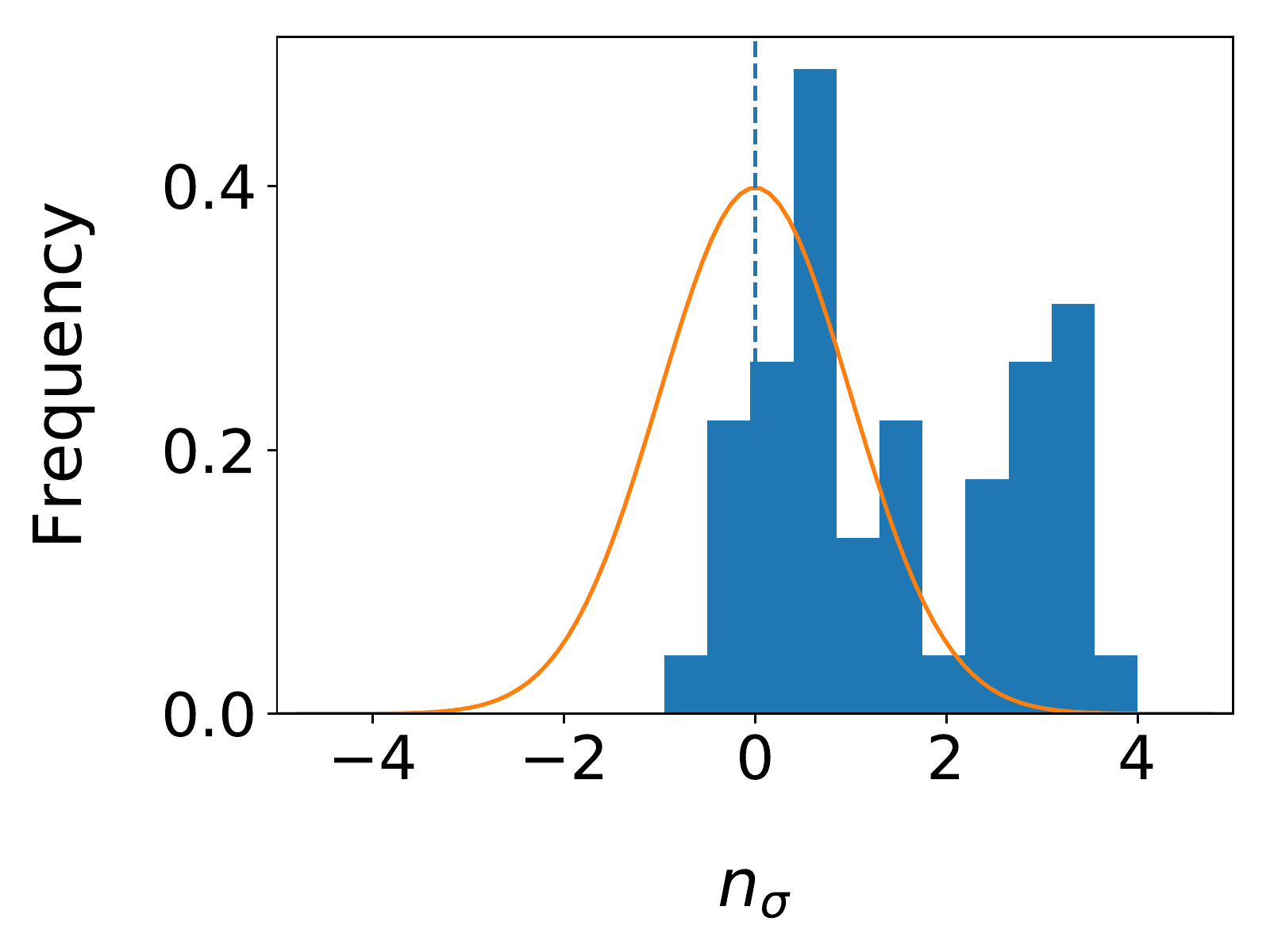}
    }
    \caption{Plots showing the distribution of extracted Hubble constant measurements around the \citet{2018arXiv180706209P} value ($H_0 = 67.4 \pm 0.5$ km s$^{-1}$Mpc$^{-1}$), in units of quoted uncertainty, given by Equation~\ref{eq:nsigma}. Error asymmetry has been taken into account for these plots. Separate plots are shown for all extracted datapoints \protect\subref{fig:allsigma}, and the distributions of values before \protect\subref{fig:preplancksigma} and after \protect\subref{fig:postplancksigma} the 2013 Planck publication \citep[][a notable point in the recent history of the Hubble constant]{Planck+2014A&A...571A..16P}. A normal ($\mu=0$, $\sigma=1$) distribution has been overlayed for readability. The tension in the measured values of the Hubble constant may be easily discerned in these plots by the peak at approximately ${+}3.5\sigma$, which corresponds to the measurements at ${\sim}73$ km s$^{-1}$Mpc$^{-1}$, which is most strongly observed post-2013 Planck.}\label{fig:nsigma}
\end{figure*}

Clusters of values given without uncertainties may be seen at 50, 65, 70, 75, and 100 km s$^{-1}$Mpc$^{-1}$. These correspond to commonly used assumed values of the Hubble constant in cosmological simulation and approximate calculations. It is interesting to note that the usage of all but the 70 km s$^{-1}$Mpc$^{-1}$ value drops off after ${\sim}2005$, whereas the 70 km s$^{-1}$Mpc$^{-1}$ value is in use until ${\sim}2009$. These decreases seem to follow the publications of HST and WMAP, respectively, by a year or two, and it may be that the growth in popularity of the values reported by those groups may have led to a shift in any presumed value of the Hubble constant.

We may also see the spread of values decreasing with time -- both for the novel reported values, and the presumed values as mentioned above. This decrease in spread is reflected in the decrease in uncertainty on each individual measurement. These effects are to be expected, due to improvements in experimental techniques and equipment over time. However it should be noted that the provided uncertainties do not show complete agreement between the reported values, and closer examination shows two distinct groupings of measurements in the post-Planck era (ignoring a grouping at 75 km s$^{-1}$Mpc$^{-1}$, which are without uncertainties and therefore likely assumed values rather than reported), at ${\sim}68$ and ${\sim}73$ km s$^{-1}$Mpc$^{-1}$. This corresponds to a known debate in the literature, arising from the difference between the values from local measurements of the Hubble parameter \citep{Riess+2018ApJ...861..126R}, and measurements inferred from the Cosmic Microwave Background \citep{Planck+2014A&A...571A..16P}, where the former finds a value of $67.4\pm 0.5$ km s$^{-1}$Mpc$^{-1}$ and the latter $73.45\pm 1.66$ km s$^{-1}$Mpc$^{-1}$ -- a $3.5\sigma$ discrepancy. This tension may be due to uncorrected systematic errors in the data, new physics, or an unknown feature of one or both data sets, and each of these possibilities has been debated in the literature \citep{t1,t2,t3,t4,t5,t6,t7,t8,t9}.

To better illustrate this discrepancy, the distribution of extracted values has been plotted in reference to the \citet{2018arXiv180706209P} value of the Hubble constant ($H_0 = 67.4 \pm 0.5$ km s$^{-1}$Mpc$^{-1}$), in units of quoted uncertainty (see Figure~\ref{fig:nsigma}). Following \citet{Croft+Dailey2011arXiv1112.3108C}, all extracted measurements which include an uncertainty have been converted into a $\sigma$ difference from this reference value, according to,
\begin{equation}\label{eq:nsigma}
    n_{\sigma}=(H_{0,\textrm{measured}} - H_{0,\textrm{true}})/\sigma_{\textrm{measured}},
\end{equation}
where $H_{0, \textrm{true}}$ is the aforementioned reference value, and $H_{0,\textrm{measured}}$ and $\sigma_{\textrm{measured}}$ are the extracted value and uncertainty. Asymmetric uncertainties have also been accounted for. We may clearly see in Figure~\ref{fig:postplancksigma} (showing measurements published after \citealt{Planck+2014A&A...571A..16P}) a peak at approximately ${+}3.5\sigma$, corresponding to the local measurements of the Hubble constant.
This shows that our algorithm has successfully recovered the current tension in the field, and has the potential to provide an objective quantification of the consensus  of a given measureable property, and whether any tension exists within the literature.

In Figure~\ref{fig:splithist} we may also see that measurements without uncertainties are predicted to be less likely to originate in papers which are not reporting a new measurement, using our neural network from Section~\ref{sec:classifier}. This would agree with the assumption that these assumed values are primarily used in simulations, or theoretical work. It also agrees with the assumption that astrophysical articles which have a numerical value with an associated uncertainty in their abstract are likely reporting said value. It should be noted that the predictions from the ``new measurement'' classifier are not on a per-measurement basis, but rather a per-publication basis, and it is possible that a given publication will refer to both an assumed or historical value, and a novel value (with uncertainty) in the same abstract. This could account for the high positive prediction probability of some unlikely values. It should also be noted that some outlier values (for example the value at 44 km s$^{-1}$Mpc$^{-1}$ in \citet{2007MNRAS.380..669C}) are noted as such by the paper authors, who point out the inconsistency and suggest further study of the discrepancy -- nonetheless these are ``valid'' measurements from the perspective of our model, and hence their inclusion is a feature of the unbiased nature of this model.

Finally, we may see from the histogram of measurement values that there is a distinct peak in the distribution around ${\sim}70$ km s$^{-1}$Mpc$^{-1}$, which agrees with accepted wisdom on the value of the Hubble constant. However, it is noted that little structure is apparent in the time-domain histogram. There appears to be an increase in the number of publications reporting a new value of the Hubble constant in the months preceding the publication of WMAP, but this same trend is not clear for the other landmark publications -- and the dearth of publications following WMAP is, perhaps, puzzling.

\section{Conclusions} \label{sec:conclusion}

We present, to the best of our knowledge, the first attempt to automate the extraction of measured values from the astrophysical literature, using the Hubble constant for our pilot study. Our model has successfully extracted measurements of the Hubble constant from a corpus of 208,541 arXiv astrophysics papers, published between July~1991 and September~2017, finding 573 measurements from 477 papers. We demonstrate that the rules-based model, a classical technique in natural language processing, is a powerful method for extracting measurements of the Hubble constant from a large number of publications.
We have also developed an artificial neural network model to identify papers which report novel measurements. The model was trained using article abstracts as input data with the training data taken from our ``silver'' dataset, which was constructed using information present in article titles and conclusions. We applied the neural network model to the available arXiv data, and demonstrated that our model works well in identifying papers which are reporting new measurements. From the analysis of our results we find that reporting measurements with uncertainties and the correct units is critical information to identify measurements in free text.

Our results correctly highlight the current well-known tension for measurements of the Hubble constant. This demonstrates that the tool presented in this paper is useful for meta-studies of astrophysical measurements, and shows the potential to generalise this technique to other areas.

However, in its current form the algorithms presented in Section~\ref{sec:measurementextraction} have some limitations. We are able to extract measurements of entities with a small set of simple, atomic names -- i.e. where there is a set of continuous strings, each with little or no variation (e.g. capitalisation). This is ideal for entities such as the Hubble constant, which has only a handful of standard linguistic and mathematical expressions (listed in Section~\ref{sec:measurementextraction}), and can therefore be easily encoded for searching free text.
However, the use of regular expressions and simple keyword searches make this system fragile against minor variations in standard syntax and typesetting, which is hard to account for manually. Additionally, if we consider a more complex entity (from a linguistic standpoint), such as ``the radius of the Milky Way'', we may imagine many constructions of this in written English, followed by the problem of the lack of a standardised mathematical symbol for this quantity. Our algorithm is currently unable to deal with such linguistic complexity without a large amount of effort on the part of the user to list the many possible variations of an entity's name -- and, indeed, this would also lead to the problem that the user may be unaware of many common constructions of the entity they are searching for, which will lead to poor recall.

Further, there are difficulties associated with our algorithm's assumption that all measurements appear in the same sentence as the name of the entity to which they belong. This is problematic as an assumption for two primary reasons: First, most simply, there are instances where this assumption is broken. This can occur due to complex or convoluted sentence construction, or the presence of many caveats and contingent information. A second, more involved problem is the circumstance where a measured entity has no agreed upon mathematical symbol, and one is assigned to it earlier in the text -- or where there is an agreed-upon symbol, but it is commonly used elsewhere (e.g. $\mu$) and hence is defined for the reader. In such a scenario the user can only reasonably supply a written name for the quantity they are searching for, but in many cases we may find the final result reported using its locally-agreed-upon symbol. In its current form the model cannot account for this kind of relationship.



The next stages for this project shall involve the use of more advanced natural language processing techniques to solve these problems. We shall explore the use of traditional information extraction approaches and modern neural techniques to improve the versatility of the search algorithms with respect to entity names and relationships above the sentence-level. Further, we will experiment with named-entity extraction techniques to automatically detect parameter names, allowing for the creation of a database of named measurements without the need for human specified entity names. As part of this we shall be exploring relationships within complex entity names. This would, for example, allow for automatically detecting that the named entities, ``mass of the Milky Way'', and, ``radius of the Milky Way'', are both statements regarding properties (mass and radius) of the same object (the Milky Way). This would allow for more sophisticated database population, and therefore greater utility for the user. Future work will also deal with expanding the scope of the model to include the extraction of contingent information, such as experimental technique and stated parameters (such as assumed cosmology in cosmological simulation papers).


\section*{Acknowledgements}

This work was supported by an Allen Distinguished Investigator Award, the Science \& Technology Facilities Council (STFC Grant ST/N000811/1), and the STFC Centre for Doctoral Training in Data Intensive Science, UCL. Thomas D. Kitching is supported by a Royal Society University Research Fellowship.





\bibliographystyle{mnras}
\bibliography{ms} 






\bsp	
\label{lastpage}
\end{document}